# Self-induced Magnetic Flux Structure in the Magnetic Superconductor RbEuFe$_4$As$_4$

V. K. Vlasko-Vlasov[1], A. E. Koshelev[1], M. Smylie[1,2], J.-K. Bao[1], D. Y. Chung[1], M. G. Kanatzidis[1,3], U. Welp[1], and W.-K. Kwok[1]

[1] Materials Sciences Division, Argonne National Laboratory, Argonne, Illinois 60439
[2] Department of Physics and Astronomy, Hofstra University, Hempstead, New York 11549
[3] Department of Chemistry, Northwestern University, Evanston, Illinois 60208

We report an unusual enhancement of the magnetic induction in single crystals of the magnetic superconductor *RbEuFe$_4$As$_4$*, highlighting the interplay between superconducting and magnetic subsystems in this material. Contrary to the conventional Meissner expulsion of magnetic flux below the superconducting transition temperature, we observe a substantial boost of the magnetic flux density upon approaching the magnetic transition temperature, $T_m$. Direct imaging of the flux evolution with a magneto-optical technique, shows that the magnetic subsystem serves as an internal magnetic flux pump, drawing Abrikosov vortices from the surface, while the superconducting subsystem controls their conveyance into the bulk of the magnetic superconductor via a peculiar self-organized critical state.

The co-existence of superconductivity and magnetism in the new family of rare-earth iron pnictides with high $T_c$ and comparable magnetic Curie points [1-7], provides a rare glimpse into the interplay of these typically antithetical phases. Compared to the low-temperature magnetic superconductors, where weak magnetism accompany superconductivity only in a very narrow temperature window [8,9], in the new pnictides the superconducting and magnetic orders robustly coexist over a wide range of temperatures. For example, in phosphorus-doped *EuFe$_2$As$_2$*, the superconductivity appears below $T_c$~24K, while at the Curie point, $T_m$~19K, the *Eu*-spins order ferromagnetically in layers separated by the superconducting *FeAs*-sheets and align parallel to the *c*-axis [10]. Recent MFM studies have found that in this material the Meissner state coexists with very fine ferromagnetic domains that are smaller than the penetration depth and at lower temperatures transforms into a vortex-domain state [11,12].

In this Letter, we report observation of an intriguing magnetic flux behavior in the latest member of the stoichiometric iron-pnictide superconductor family, *RbEuFe$_4$As$_4$*, with a superconducting transition at $T_c \sim 37$ K and the onset of long-range magnetic order of the *Eu* moments near $T_m \sim 15$ K [3-5,13]. In this material, the *Eu*-moments order ferromagnetically within the *ab*-plane and have the in-plane orientation. Even though the *Eu*-layers are separated by two superconducting *FeAs*-sheets, non-negligible antiferromagnetic exchange interactions occur along the *c*-axis. Resonant elastic x-ray scattering data revealed modulations along the *c*-axis with a period of 4 lattice parameters [14]. This corresponds to the helical ordering of $Eu^{2+}$ spin layers coupled by a weak antiferromagnetic exchange along the *c*-direction (see e.g. [15]). Previous work has shown [13] that this magnetic state is rather fragile and is easily polarized in the *ab*-plane by modest fields of less than 1 kG. Temperature and field dependences of the specific heat point to Berzinskii-Kosterlitz-Thouless nature of the magnetic transition at $T_m$, as described by two-dimensional anisotropic Heisenberg model, with fine features caused by the three-dimensional effects [16].

These remarkable materials characteristics afford access to the interplay between superconductivity and magnetism that is not easily realized in other magnetic superconductors. Using magnetization measurements and direct magneto-optical imaging of the distribution of magnetic flux, we discovered that at $T \lesssim T_m$, *Eu* magnetic order enhances the magnetic induction and induces a self-generated critical state, which in turn is balanced by superconducting currents. The images reveal a highly non-uniform flux distribution with high concentrations near the edges of the sample, which we ascribe to anisotropic pinning and to feedback between magnetization and field-dependent critical currents. The vortex dynamics associated with the self-generated critical state induces magnetization curves that resemble an apparent paramagnetic Meissner effect.

Single crystals of $RbEuFe_4As_4$ were grown by flux method [17]. Samples in the shape of rectangular platelets, few hundred micron lateral and few tens of micron thick, were previously characterized by magnetization, transport, specific heat, and X-ray techniques [13]. They revealed a sharp superconducting transition at $T_c$~37K with $\Delta T$~0. 5K and a magnetic ordering transition at $T_m$~15K.

We used SQUID magnetometry to characterize the temperature variation of the macroscopic magnetic moments, $M(T)$, during cooling and warming cycles in constant magnetic fields parallel to the *ab*-plane. The magnetic flux distribution inside the samples during temperature cycling was obtained with the magneto-optic imaging (MOI) technique [18]. The edge face of several platelet crystals was polished perpendicular to the *ab*-plane after gluing the samples between two aluminum blocks. The assembly was placed in an optical cryostat, covered with a MOI indicator film, and imaged with a polarized light microscope. In field-cooling (FC) experiments, the magnetic field was applied perpendicular to the polished sample face (***H***// *ab*-plane) at temperatures $T>T_c$ and MOI images of the magnetic flux distribution within the samples were recorded with decreasing $T$ down to 5K and then with gradually increasing $T$ back to above $T_c$. In zero-field-cooling (ZFC) experiments, the samples were cooled to 5K followed by field application, and flux images were recorded upon increasing $T$.

The macroscopic magnetic response during field-cooling in different magnetic fields ***H**_a* ∥ *ab*-plane is shown in Fig. 1a. At small $H_a$, the diamagnetic signal at $T_c$~37K is very weak and is followed by a rapid increase in magnetization at $T_m$ ~15K and saturation at $T<T_m$. For fields larger than 10 Oe, the $M(T)$ curves show clear paramagnetic response ($M>0$ at all temperatures) that increases with $H_a$. At $T_c$, the FC curves have a small downward kink (inset in Fig.1a) revealing the superconducting contribution.

The ZFC temperature dependent magnetization curves in Fig.1b, measured during field-warming, show mostly diamagnetic behavior at low temperatures and small $H_a$, and a $M(T)$ maximum around $T_m$. $M(T)$ crosses the $M=0$ line for $H_a >150$ Oe and the positive magnetization segment expands with further increasing field. This agrees with the behavior of $M(H)$ loops for $\boldsymbol{H}\|ab$ [13], which depict the positive growth of $M$ at $T<T_m$ due to the twist of $Eu^{2+}$ spins towards $\boldsymbol{H}$. The $M(H)$ loops for $\boldsymbol{H}\|c$-axis [13] are strongly tilted showing a large magnetic anisotropy, which locks the $Eu^{2+}$ moments in the $ab$-plane.

*FC flux patterns*. The temperature variation in the MOI contrast around the samples is hardly visible during FC in small $H_a$. However, distinct features corresponding to the appearance of the ferromagnetic response in the macroscopic $M(H)$ curves emerge in the MOI images at $H_a \gtrsim 50$ Oe. Figure 2 shows a set of images for one of the samples cooled in a field of 220 Oe. Above $T_c$, the contrast is homogeneous (Fig.2a) confirming that the sample's magnetization at $T \geq T_c$ is small and does not perturb the uniformity of the applied field. With decreasing temperature, at $T \lesssim 30$K a bright contrast revealing the increased magnetic induction, $B>H_a$, appears around the sample (Fig.2b). This contrast, corresponding to the paramagnetic sample response, increases gradually upon further cooling (Fig.2c). Near the magnetic transition, $T_m \sim 15$K, there is a sharp increase in the contrast at the narrow edges of the sample associated with considerably enhanced $B$ (Fig.2d), which expands towards the interior of the crystal forming narrow channels along the crystal midsection. At lower $T$, the contrast changes more gradually and saturates (Fig.2e-f). A detailed movie of the magnetic flux evolution during field-cooling is presented in the supplementary material [19].

Subsequent warming of the sample after cooling to 5K, reproduces the described flux patterns in the reverse order, although with some temperature hysteresis. Similar hysteretic behavior is observed in the macroscopic $M(T)$ curves measured during Field-Cooling-Warming cycles at $H>50$ Oe (see $H_a=500$ Oe curves in Fig.1a).

At higher applied fields, the above scenario recurs with bright contrast emerging at higher temperatures, the enhanced flux regions penetrating deeper into the bulk, and larger maximum induction forming near the sample boundaries. Patterns similar to those in Fig.2 but with a different position of the narrow flux penetration channels were observed in all studied samples. A flux profile, $B(x)$, across the sample illustrated in Fig.2 field-cooled to $T=7K$ in $H_a=330$ Oe, is presented in Fig.3a. It shows that at the long edges, the flux density is enhanced by ~15% compared to $H_a$. At the narrow edges the enhancement is an order of magnitude larger.

In Fig.3b we present temperature variations of the flux profile along the midsection for another sample measured during field cooling in $H_a=440$ Oe. The induction shows sharp peaks near the narrow sample edges that increase rapidly as $T$ approaches the magnetic transition point, $T_m$. At $T<T_m$ the $B(x)$ profile changes only slightly and the maximum $B$ saturates. It reaches a value more than twice larger than $H_a$, highlighting the strong enhancement of the sample magnetization upon transition into the ferromagnetic state.

*ZFC flux patterns*. Induction patterns observed under ZFC conditions, are presented in Fig.4 showing characteristic changes of the flux distribution upon warming the sample in a field of $H_a=330$ Oe applied at 5K. The fixed applied field appears as a bright contrast around the sample perimeter which does not vary with $T$. The dark contrast over the major portion of the crystal signals the screening of the field in the bulk of the sample. At $T=5K$ (Fig.4a) the flux enters from the narrow sample edges and forms narrow channels that penetrate into the bulk, similar to those observed in the FC case at $T<T_m$. At these edges, $B$ is noticeably larger than $H_a$ already at 5K. There is also some enhancement of $B$ along the long sides of the sample. With increasing $T$, the enhanced flux regions expand deeper into the bulk upon approaching $T_m$ (Fig.4b). However, at $T>$

$T_m$ the induction in the flux occupied areas rapidly decreases, the flux disperses from the regions of maximum $B$ into surrounding areas and exits the sample (Fig.4c). As the temperature approaches $T_c$ the flux spreads even further into the bulk yielding the smooth light contrast over the entire sample (Fig. 4d). Meanwhile, the average induction in the sample remains smaller than $H_a$. This corresponds to the diamagnetic contribution to the macroscopic magnetization near $T_c$.

The described changes in the induction patterns is a specific property of *RbEuFe$_4$As$_4$* that is not observed in typical superconducting materials. The FC patterns shown in Fig.2 can be ascribed to a peculiar critical state generated by enhanced magnetic flux density and anisotropic vortex pinning. Upon approaching $T_m$, the magnetic subsystem tends to increase $M$ due to increased susceptibility which results in an increase in $B$. In the superconducting state, $B$ in the bulk can only be increased by entry of vortices from the sample surface. The vortices penetrate the sample against the pinning force causing the decay of $B$ with distance $x$ from the surface. At $T>T_m$, the magnetic susceptibility $\chi$ and the critical current $J_c$ representing the vortex pinning are small, so that the enhancement of $B$ near the sample surface and the slope of the decaying $B(x)$ are small. With further cooling towards $T_m$, both $\chi$ and $J_c$ increase, resulting in a higher $B(x)$ slope near the surface. Hence as the sample is cooled, the flux profile changes from being shallow in the bulk to sharp near the sample edges. As a result, the induction forms a *nonlinear* critical state profile *B(x)*. In our case, due to the vortex pinning anisotropy arising from the layered crystal structure, the vortices penetrate preferentially from the narrow sample edges along the ab-planes, while their entry from the large ab-surfaces is delayed. This explains the appearance of the enhanced magnetic flux near the narrow sample ends in Fig.2 as $T$ approaches $T_m$. The anisotropic $J_c$ resulting from the pinning anisotropy could explain the tapered flux penetration channels that may appear due to current flow instability in strongly anisotropic superconductors [20]. Flux channels can also result from the compositional variation in the layered crystal structure, typical for such crystals grown by the flux method. Although the Energy Dispersive X-ray Spectrometry (EDX) test of our samples did not reveal such variations within +/-3% accuracy, a smaller compositional change could yield a weak-link channel for vortex entry. Qualitative analysis of the critical current from the *M(H)* loops shows that the critical current (i.e. pinning) drops rapidly at relatively small fields (see Fig.1S in [21]). Such a fast decay of $J_c(H)$ can provide a positive feedback resulting in advanced flux penetration around the weak channels. In this case, at a sufficient density of entering vortices the field suppressed pinning will allow their additional entry and thus enhance local *B* and extend the initial flux penetration region (see Fig.2S in [21]).

In all cases, during field-cooling, at $T > T_m$, *B* is slightly enhanced near the sample surface but does not form penetration channels. However, in the vicinity of $T_m$ the flux density inside the magnetic superconductor becomes much larger than the applied magnetic field. The magnetic flux concentrates near the sample boundaries and decays towards the sample interior.

At low *T,* vortex pinning which increases with cooling [21], prevents further penetration of vortices and the average induction in the flux occupied areas saturates. Subsequent warming of the sample reduces $\chi$ and triggers the partial exit of vortices, which is delayed due to pinning. This results in the hysteresis, which we observe in the flux patterns and in the macroscopic *M(T)* curves during thermal cycling. The same behavior becomes more pronounced at higher $H_a$, where the entering flux is larger and penetrates deeper into the sample. Similar features of the FC flux evolution were observed in several samples with different width-to-thickness ratios.

The described flux distributions are in striking contrast to the FC patterns in regular superconducting plates in parallel fields. In the latter, the flux remains at the level of $H_a$ in the interior of the sample and decays to a minimum at the sample surface. Consequently, the FC induction profile *B(x)* across the SC plate has an inverted letter-M shape following the temperature evolution of $J_c$ (see Fig.5a). This profile corresponds to the diamagnetic surface Meissner current $J_M$ precipitating the drop of *B* within the penetration depth $\lambda$, and smaller critical currents $J_c$ circulating in the opposite (paramagnetic) direction, restricting the exit of vortices. In our *RbEuFe$_4$As$_4$* crystals, the FC *B(x)* profiles acquire a direct letter-M shape (Fig.3), which could be naively described by the above current pattern but with opposite current chirality (Fig.5b).

A more plausible construction of *B(x)* consists of a strong enhancement of $B=\mu(T)H_a$ (with $\mu=1+4\pi\chi \gg 1$ at $T \lesssim T_m$) in a thin surface layer and a diamagnetic $J_M$ that induces a sharp drop of *B* in the $\lambda$-layer, followed by a slow decay of induction due to the diamagnetic critical current $J_c$, which limits the additional flux penetration (Fig.5c). In the middle of the sample, screened from the entry of new vortices, the induction is trapped at the level of $B(T_c)= \mu(T_c)H_a$. The nonlinear *B(x)* decaying towards the center of the sample will follow the variation of $J_c(T)$ during cooling. In an infinitely wide plate perpendicular to *x*, if we admit that $\mu$ changes with temperature much faster than $J_c$ and assume the linear magnetic response $M=(\chi/\mu)B$, the induction profile *B(x)* across the sample, following from $\frac{dB}{dx} = \frac{4\pi}{c}J_c(T) + 4\pi\frac{dM}{dx}$, will be:

$$B(x) = \mu(T)H - \Delta B_M - \frac{4\pi}{c}\mu(T)J_c(T)x \quad (1)$$

Here $\Delta B_M$ is the surface step due to the diamagnetic $J_M$. Note, that unlike in a regular superconductor, the linear decay of B(x) is defined not by $J_c$ but by $\mu J_c$ which can be regarded as an apparent enhancement of pinning. However, flux pinning and hence $J_c$ are not enhanced by the factor of $\mu$ and only the gradient of *B(x)* increases. Within the above approximations, the depth of the vortex penetration $x_c$ following from the condition $B(x_c)= \mu(T_c)H$ can be estimated as:

$$x_c = \frac{[\mu(T)-\mu(T_c)]H - \Delta B_M}{\frac{4\pi}{c}\mu(T)J_c(T)} \qquad (2)$$

This simple model gives a rough estimate of the effect of the magnetic subsystem on the flux entry in the magnetic superconductor.

In the ZFC case, the sample cooled with $H_a=0$ is devoid of vortices. When the field is applied at low $T$, the induction is enhanced in the sample surface layer due to the large $\chi$ but the flux does not propagate far inside due to the high $J_c$. With increasing $T$, the critical current decreases and vortices enter and move towards the interior, forming flux patterns similar to those observed in the FC case. At $T>T_m$, the susceptibility drops and the average flux density inside the sample rapidly reduces as vortices disperse from the maximum $B$ areas. The reduction of $J_c$ with temperature allows the flux to redistribute over the sample and partially exit along the *ab*-planes across the narrow edges. Both the decrease in $\chi$ and $J_c$ yield the observed decay of the macroscopic *M(T)* at $T>T_m$.

Our results also provide a glimpse into an interesting scenario considered by Tachiki et al. [22,23] which can account for a large enhancement of *B* at the boundaries of ferromagnetic superconductors. Due to nonlocal magnetic response, in the vicinity of $T_m$ the persistent $J_M$ can form interleaved dia- and paramagnetic layers near the surface, which results in the paramagnetic divergence of *B* in the $\lambda$-layer. In the pure London approximation (the coherence length $\xi=0$) the oscillating $J_M$ on the very surface remains diamagnetic, but at a finite $\lambda/\xi$ the surface $J_M$ reduces to zero, which expands the enhanced *B* region [23]. Although the calculations in [22,23] use a specific set of material parameters and do not account for vortex entry, the $J_M$ oscillations could cause the surface enhancement of *B* in our crystals near $T_m$.

We note that the magnetic response illustrated in Figs.1-4 differs essentially from the paramagnetic Meissner effect in non-magnetic high-$T_c$ and some low-$T_c$ superconductors (see review [24] ), where the positive $M$ signal was found in FC-curves only at very small fields and decreased with $H_a$. In these cases, the weak paramagnetic component, which was associated with Andreev bound states in $d$-wave superconductors [25] or with flux compression effects in low-$T_c$ samples [26], was easily dominated by the regular diamagnetic response at larger $H_a$. In contrast, we find robust enhancement of positive $M$ ($\parallel H_a$) and $B$ with field, confirming the concerted action of the supercurrents and magnetic spin system in *RbEuFe$_4$As$_4$*.

In conclusion, we discovered a unique vortex matter behavior driven by the interplay between magnetism and superconductivity in the novel high-temperature magnetic superconductor, *RbEuFe$_4$As$_4$*. The magnetic flux distributions imaged with the magneto-optical technique during field-cooling and warming of the samples are consistent with macroscopic magnetic measurements and reveal a noticeable enhancement of the magnetic induction in the sample due to self-generated vortex entry at temperatures near the magnetic transition point, $T_m$. The enhancement of $B$ is especially strong at $T_m$ and progressively increases with field in the studied field range. This evolution of $B$, unlike the flux expulsion observed in traditional superconductors, results from the collective response of the magnetic and superconducting subsystems in *RbEuFe$_4$As$_4$*. Here, the magnetic subsystem acts as an absorption pump, drawing magnetic flux in when the magnetic susceptibility increases in the vicinity of $T_m$, while the superconducting subsystem provides the delivery of magnetic flux into the sample through the generation of Abrikosov vortices at the sample surface and their pinning-controlled penetration into the bulk. The enhancement of $B$ inside the samples, produced by the entry of additional vortices and their propagation to distances limited by pinning, is strongly anisotropic in the layered *1144* crystals and yields peculiar inhomogeneous vortex density patterns. Together, they reveal the unique self-organization of magnetic flux dynamics in magnetic superconductor *RbEuFe$_4$As$_4$*.

This work was supported by the U.S. Department of Energy, Office of Science, Materials Sciences and Engineering Division.

*Figures*

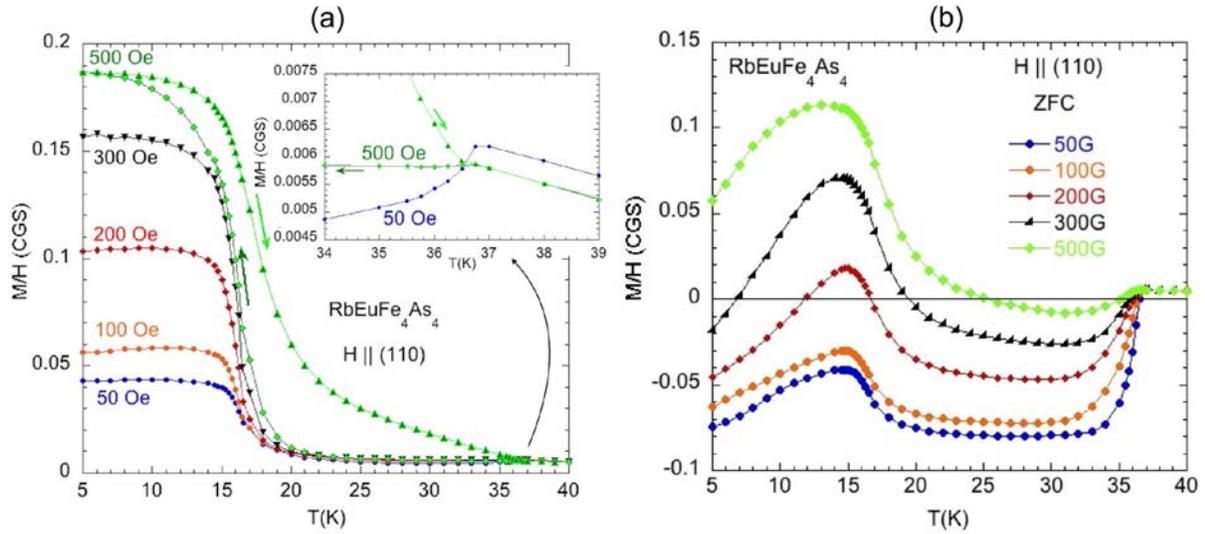

Fig. 1 (a) Field-cooled (FC) and (b) zero-field-cooled (ZFC) *M(T)* curves at different fields $H_a$|| *ab*-plane. In (a) the field-cooling-warming hysteresis is shown for $H_a$=500 Oe. The insert in (a) presents the expanded region of the FC curves near $T_c$. The magnetization is referenced to $H_a$.

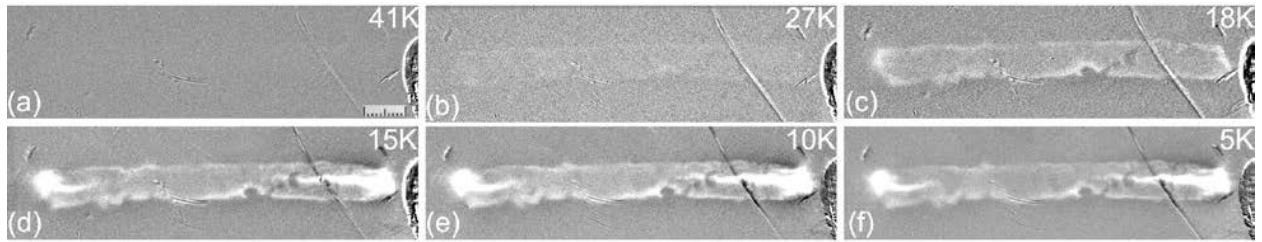

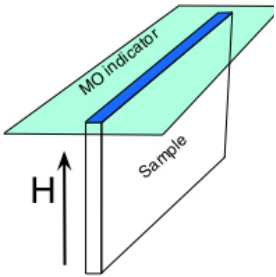

Fig. 2 a-f Magneto-optical images of the flux evolution in a *RbEuFe₄As₄* crystal during field-cooling in $H_a$=220 Oe. The experimental geometry is shown in (g). The contrast brightness corresponds to the strength of $B$||$H$. Scale bar in (a) is 100 μm.

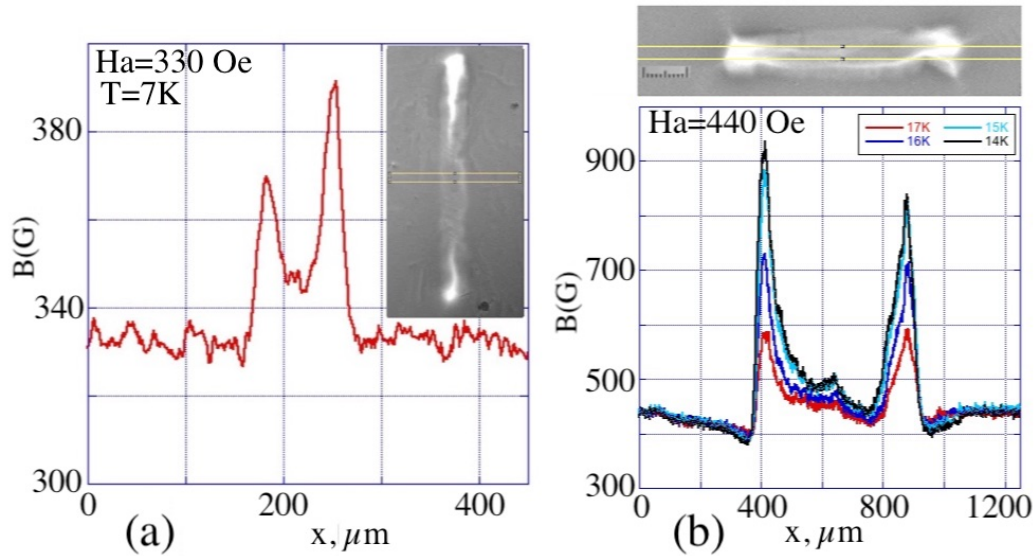

Fig.3 (a) Induction profile $B(x)$ across the sample shown in Fig.2 after field-cooling to 7K in $H_a$=330 Oe. (b) Temperature evolution of the $B$ profile along another sample during field-cooling in $H_a$=440 Oe. $B$ is averaged over the width of the bands shown by yellow lines in the inserts. The scale bar in the insert of (b) is 100 μm. The field is decreased outside the sample in (b) due to the stray fields of the enhanced $B$ inside the plate.

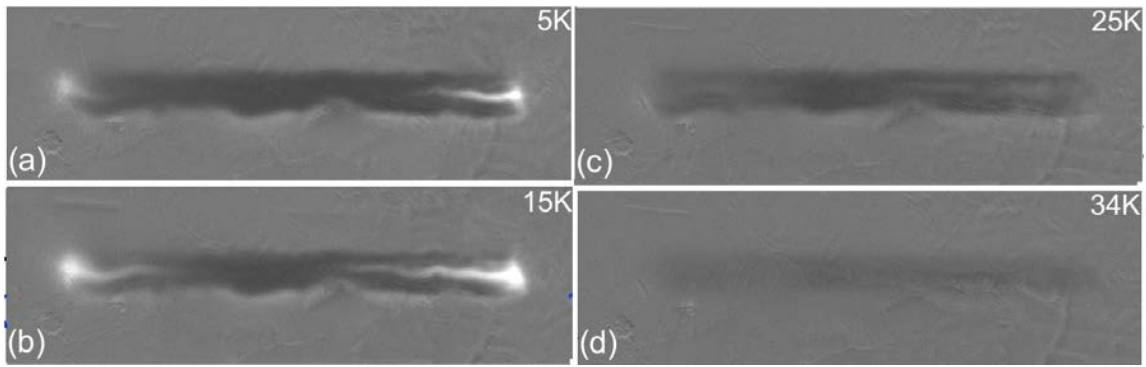

Fig. 4 Magnetic flux patterns in the sample shown in Fig.2 during warming in $H_a$=330 Oe after zero-field-cooling to 5K.

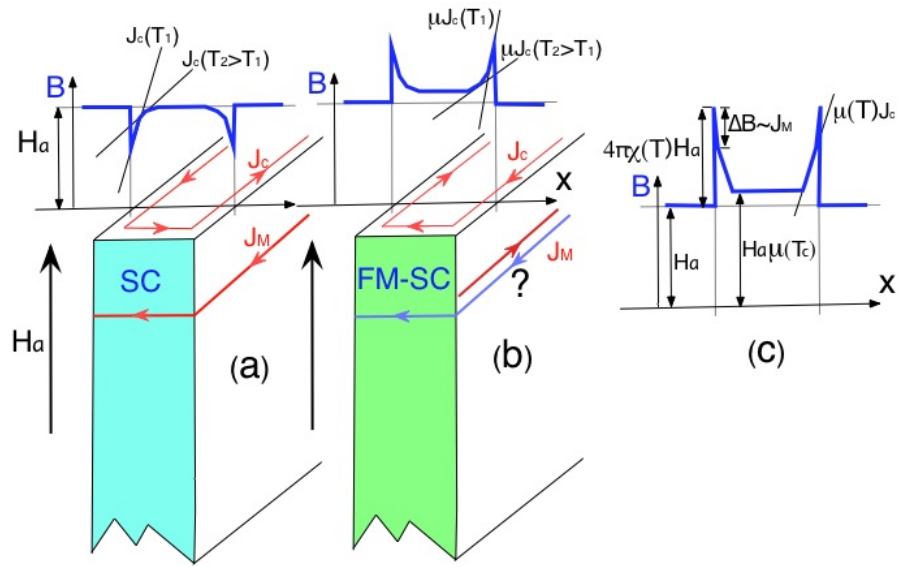

Fig. 5 Sketch of the induction distribution in a usual field-cooled superconducting plate (a) , in our ferromagnetic superconductor (b), and in FM-SC with temperature independent $J_c$ (c).

Supplemental information for
"**Self-induced Magnetic Flux Structure in Magnetic Superconductor RbEuFe4As4**"

V. K. Vlasko-Vlasov[1], A. E. Koshelev[1], M. Smylie[1,2], J.-K. Bao[1], D. Y. Chung[1], M. G. Kanatzidis[1,3], U. Welp[1], and W.- K. Kwok[1]

[1] Materials Sciences Division, Argonne National Laboratory, Argonne, Illinois 60439
[2] Department of Physics and Astronomy, Hofstra University, Hempstead, New York 11549
[3] Department of Chemistry, Northwestern University, Evanston, Illinois 60208

Fig.1S The width, $\Delta M$, of the magnetic hysteresis loops $M(H)$ measured at several temperatures above and below the magnetic transition point $T_m$=15K of *RbEuFe$_4$As$_4$* for magnetic field parallel to the *ab*-plane. The right plot shows the decay of $\Delta M$ with temperature at several fields. The sample is a single crystal plate with large surfaces parallel to *ab*-plane.

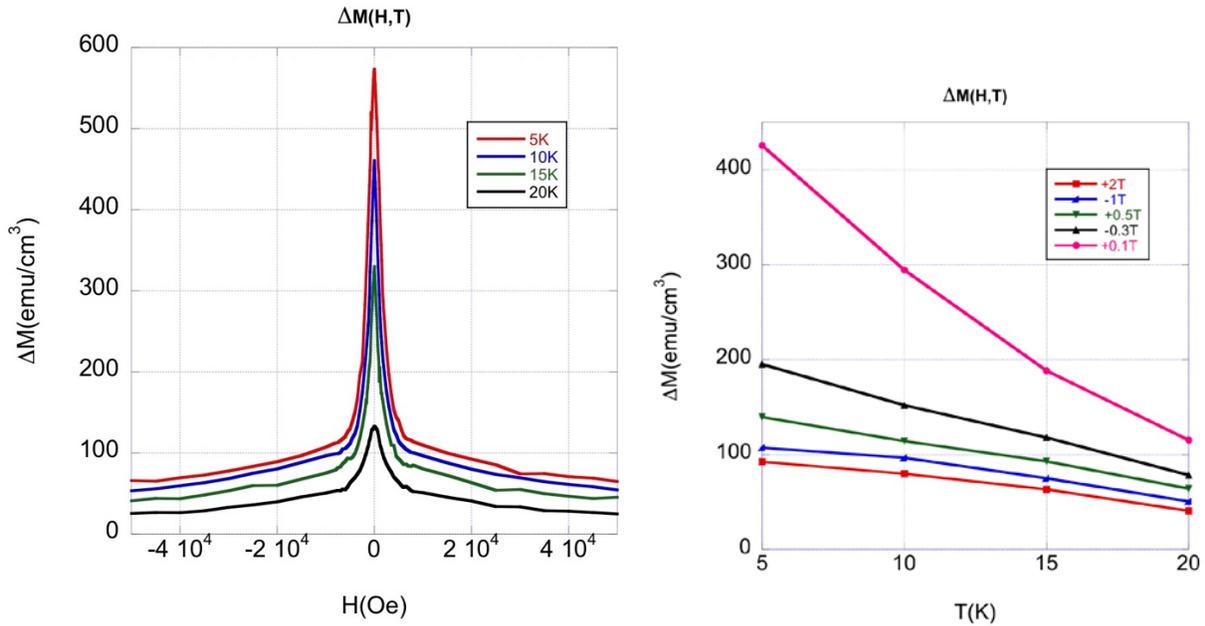

In usual type II superconductors, the hysteresis width $\Delta M$ reflects the strength of vortex pinning by defects in the sample or at the surface. Vortex pinning determines the value of the critical current $J_c$ that can be extracted from $\Delta M$ using a critical state model and accounting for the sample geometry. However, in magnetic superconductors the interplay of magnetism and superconductivity results in a complex relationship between $\Delta M$ and $J_c$ which does not allow a direct quantitative estimate of $J_c$. At the same time, the qualitative correspondence remains. For example, above plots indicate a rapid drop in $J_c$ with magnetic field and a concurrent increase of pinning and thus the $J_c$ value with decreasing temperature.

Fig.2S Sketch of the advanced vortex penetration due to feedback caused by the field suppression of $J_c$.

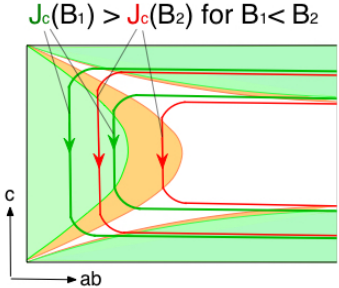

$J_c$ density corresponds to the inverse distance between the current lines. Additional vortices are generated at the surface layer by enhanced magnetic susceptibility at $T_m$ and enter preferentially along the *ab*-direction due to the vortex pinning anisotropy. The resultant increased $B$ suppresses $J_c$ (i.e. decreases pinning) and the flux entry front advances deeper into the sample (vortex phase expands from green to brown area) .